\documentclass[3p,times,procedia]{elsarticle}
\usepackage{nupha_ecrc}
\usepackage{amssymb}
\usepackage[figuresright]{rotating}
\usepackage{color}
\usepackage{ulem}

\volume{00}
\firstpage{1}
\journalname{Nuclear Physics A}
\runauth{}
\jid{nupha}
\jnltitlelogo{Nuclear Physics A}

\begin{document}

\begin{frontmatter}



\dochead{XXVIth International Conference on Ultrarelativistic Nucleus-Nucleus Collisions\\ (Quark Matter 2017)}

\title{Dynamical fluctuations in critical regime and across the 1st order phase transition }


\author[1,2]{Lijia Jiang}
\author[2]{Shanjin Wu}
\author[2,3,4] {Huichao Song}

\address[1]{ Frankfurt Institute for Advanced Studies,Ruth Moufang Strasse 1, D-60438,
Frankfurt am Main, Germany}
\address[2]{ Department of Physics and State Key Laboratory of Nuclear Physics and
Technology, Peking University, Beijing 100871, China}
\address[3]{ Collaborative Innovation Center of Quantum Matter, Beijing 100871, China}
\address[4]{ Center for High Energy Physics, Peking University, Beijing 100871, China}

\begin{abstract}
In this proceeding, we study the dynamical evolution of the sigma field within the framework of Langevin dynamics.
We find that, as the system evolves in the critical regime, the magnitudes and signs of the cumulants of sigma field, $C_{3}$ and $C_{4}$, can be dramatically different from the equilibrated ones due to the memory effects near $T_c$. For the dynamical evolution across the 1st order phase transition boundary,
the supercooling effect leads the sigma field to be widely distributed in the thermodynamical
potential, which largely enhances the cumulants $C_3, \ C_4$, correspondingly.
\end{abstract}


\end{frontmatter}


\section{Introduction}

The STAR collaboration has measured the higher
order cumulants of net protons in Au+Au collisions with collision energy
ranging from $7.7$ to $200$ GeV~\cite{Aggarwal:2010wy,Adamczyk:2013dal,Luo:2015ewa}.
The experimental data of $\kappa \sigma ^{2}\;\left( \kappa \sigma ^{2}=C_{4}/C_{2}\right) $ shows
a large deviation from the poisson baseline, and presents
an obvious non-monotonic behavior at lower collision energies,
indicating the potential of discovery the QCD critical point in experiment~\cite{Luo:2015ewa}.

Within the framework of equilibrium critical fluctuations, we calculated the fluctuations of net protons
through coupling the fluctuating sigma field with particles emitted from the freeze-out
surface of hydrodynamics~\cite{Jiang:2015hri}. Our calculations can fit the $C_4$ and $\kappa \sigma^2$ data
by tuning the related parameters, as well as qualitatively describing the acceptance dependence of
the cumulants of net protons.
However, our calculations over-predicted both $C_{2}$ and $C_{3}$ data due to the positive critical fluctuations, which are
in fact intrinsic for the traditional equilibrium critical fluctuations~\cite{Stephanov:2008qz,Stephanov:2011pb,Ling:2015yau}.

Recently,  Mukherjee and his collaborators have studied the non-equilibrium evolution for the
cumulants of sigma field in the critical regime,
based on the Fokker-Plank equation~\cite{Mukherjee:2015swa}. The numerical results  showed
that, as the system evolves near the critical points, the memory effects keep
the signs of the Skewness and Kurtosis at the early time, which are opposite to the
signs of the equilibrium ones at the freeze-out points below $T_c$.
However, their calculations focus on the zero mode of the sigma field, which has
averaged out the spatial information at the beginning and can not directly couples
with particles to compare with the measured experimental data.

To solve this problem, one could directly trace the whole space-time evolution of
the sigma field within the framework of Langevin dynamics. In this
proceeding, we will present the main results from our recent numerically
simulations of the Langevin equation of the sigma field, using an effective
potential of the linear sigma model with constituent quarks.
As discovered in early work~\cite{Mukherjee:2015swa}, we also observe
clearly memory effects as the system evolves in the critical regime,
which largely influence the signs and values of the cumulants $C_{3}$ and $C_{4}$.
In addition, we find that for the dynamical evolution across the 1st order phase transition boundary,
the supercooling effect leads the sigma field to be widely distributed in the thermodynamical
potential, which largely enhances the corresponding cumulants $C_2-C_4$ at the freeze-out
points.

\section{The formalism and set-ups}

\begin{figure*}[tbp]
\center
\includegraphics[width=2.7 in]{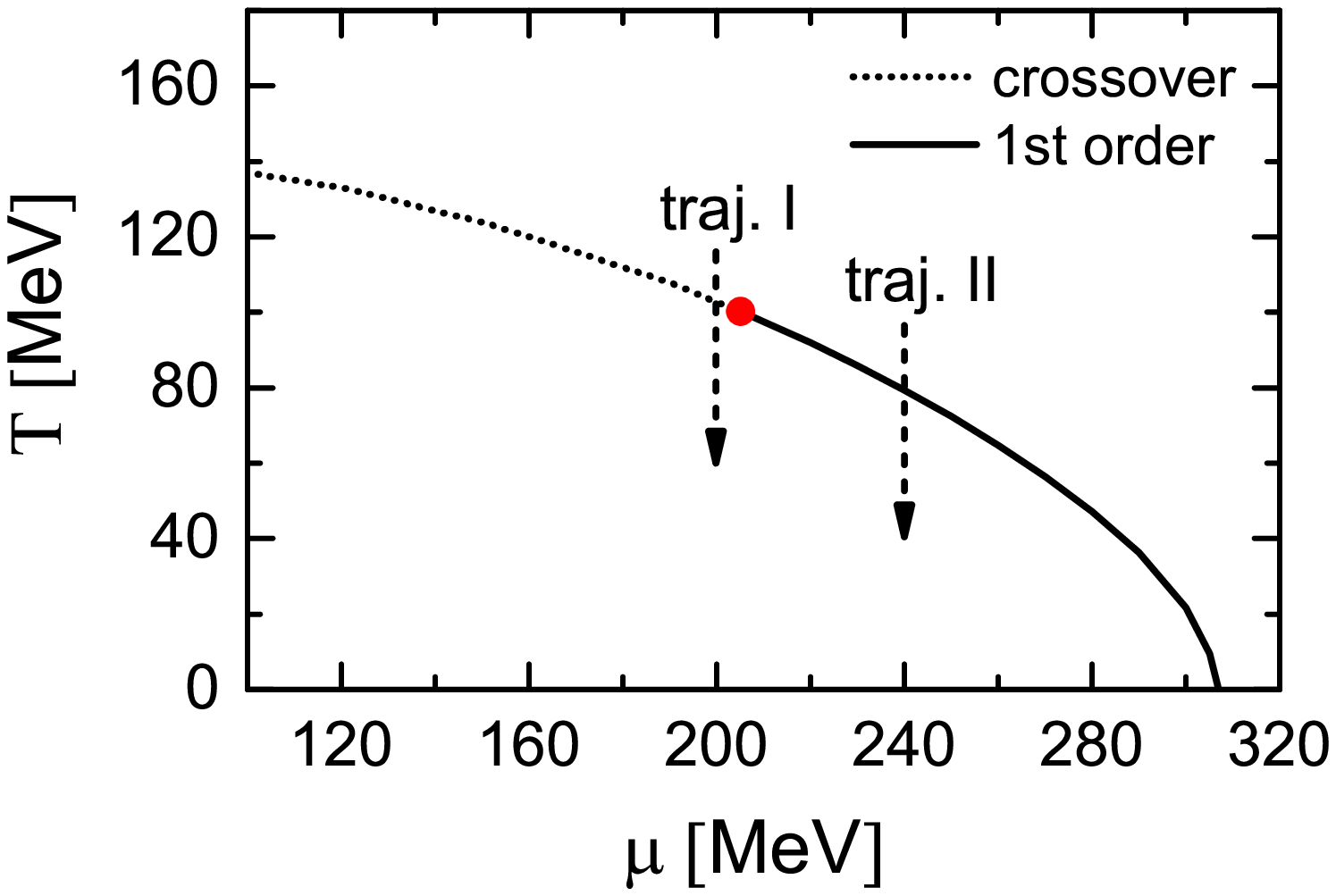}
\includegraphics[width=2.7 in]{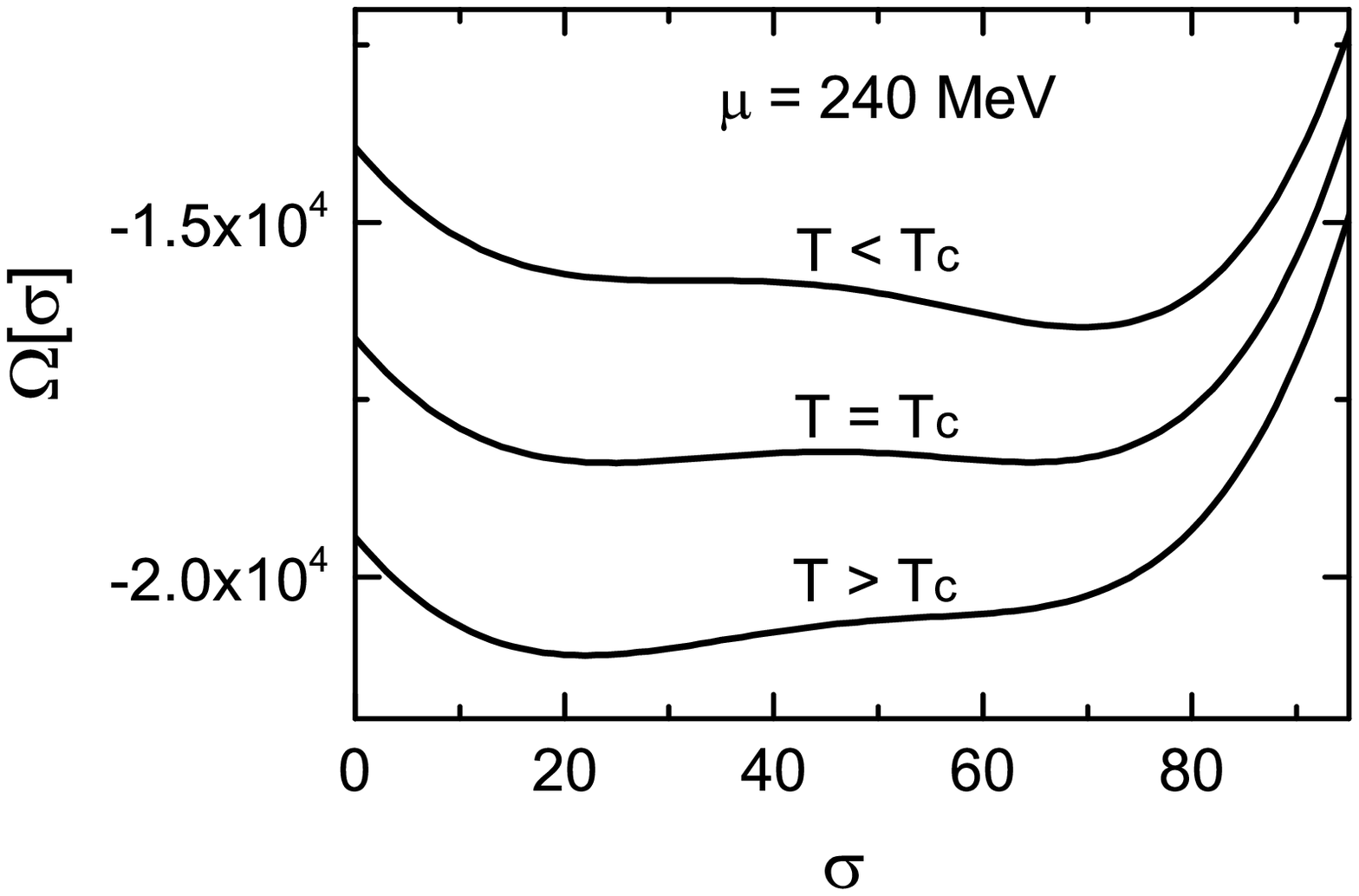}
\vspace{-7mm}
\caption{Left panel: the phase diagram on the ($T, \mu$) plane, obtained from the linear signa model with constituent quarks. Right panel: the thermodynamic potentials with different temperatures ($T<T_c$, $T=T_c$ and $T>T_c$), but with the same chemical potential $\mu = 240$ MeV.}
\label{omega}
\end{figure*}

In this proceeding we focus on the dynamical evolution of the order parameter field within
the framework of the linear sigma model with constituent quarks. According the the classification
of the dynamical universality classes~\cite{Rev1977}, our approach belongs to model A, which is not in
the same dynamical universality class of the full QCD matter evolution~\cite{Son:2004iv}, but easy for numerical
implementations.  The linear sigma model
is an effective model to describe the chiral phase transition, which
presents a complete phase diagram on the $(T,\mu )$ plane with different
phase transition scenarios, including a critical point~\cite{Jungnickel:1995fp,Skokov:2010sf}.
As the mass of the sigma field vanishes at the critical point, the related thermodynamical
quantities become  divergent due to the critical long wavelength fluctuations of the sigma
field. In the critical regime, the semi-classical evolution of the long wavelength mode of the sigma field
can be described by a Langevin equation~\cite{Nahrgang:2011mg}:
\begin{equation}
\partial ^{\mu }\partial _{\mu }\sigma \left( t,x\right) +\eta \partial
_{t}\sigma \left( t,x\right) +\frac{\delta V_{eff}\left( \sigma \right) }{%
\delta \sigma }=\xi \left( t,x\right),
\end{equation}%
where $\eta $ is the damping coefficient and $\xi \left( t,x\right) $ is the noise
term. Both of these two terms come from the interaction
between the sigma field and quarks, and satisfy the fluctuation-dissipation theorem~\cite{Nahrgang:2011mg}. Here we take $\eta$ as a free parameter, and input white noise in the calculation.  The effective potential of the sigma field is written as:
\begin{equation}
V_{eff}\left( \sigma \right) =U\left( \sigma \right) +\Omega _{\bar{q}%
q}\left( \sigma \right) = \frac{\lambda ^{2}}{4}\left( \sigma
^{2}-v^{2}\right) ^{2}-h_{q}\sigma -U_{0}+\Omega _{\bar{q}%
q}\left( \sigma \right) \qquad \quad
\end{equation}%
where $U\left( \sigma \right) $ is the vacuum potential of the chiral field, and
the related values of $\lambda$, $\sigma$, $h_q$ and $U_0$ are set by
the vacuum properties of hadrons. Note that here we have neglected the fluctuations of $\vec{\pi}$,
since its mass is finite in the critical regime. $\Omega _{\bar{q}q}$ represents
the contributions from thermal quarks, which has the form:
\begin{equation}
\Omega _{q\bar{q}}\left( \sigma ;T,\mu \right) =-d_{q}\int \frac{d^{3}p}{%
\left( 2\pi \right) ^{3}}\{E+T\ln [ 1+e^{-\left( E-\mu \right) /T}%
] +T\ln [ 1+e^{-\left( E+\mu \right) /T}] \}
\end{equation}%
where $d_{q}$ is the degeneracy factor of quarks, and the energy of the quark is $E=\sqrt{%
p^{2}+M\left( \sigma \right) ^{2}}$. Here we introduce an effective mass for the quark, $M\left( \sigma
\right) = m_0+g\sigma$~\cite{Jiang:2015hri,Stephanov:2011pb}.  After the chiral phase transition,
quarks obtain effective mass and turn to
constituent quarks. With $g=3.3$, the effective mass of the constituent quark at $T=0$ is approximately
310 MeV, which corresponds
to the proton mass in vacuum $m_p \sim 930 $ MeV. Based on the effective potential Eq.~(2), one can obtain
the corresponding phase diagram in the ($T, \mu$) plane, which is plotted in the left panel of Fig.~\ref{omega}.

For the numerical implementations, we first construct the profiles of the initial sigma field
according to the probability function  $P\left[ \sigma
\left( \mathbf{x}\right) \right] \sim \exp \left( -\varepsilon \left( \sigma
\right) /T\right) $ (where $\varepsilon \left( \sigma \right) =\int d^{3}x%
\left[ \frac{1}{2}\left( \nabla \sigma \left( x\right) \right)
^{2}+V_{eff}\left( \sigma \left( x\right) \right) \right] $),  then evolve the
sigma field event by event through solving the Langevin equation Eq.(1).  With the obtained
space-time configurations of the sigma fields, the moments of
the sigma field at a certain evolution time can be calculated as:
\begin{equation}
\mu _{n}^{\prime }=\langle \sigma ^{n}\rangle =\frac{\int d\sigma \sigma
^{n}P\left[ \sigma \right] }{\int d\sigma P\left[ \sigma \right] },
\end{equation}%
where $\sigma =\int d^{3}x\sigma \left( \mathbf{x}\right) $. The
cumulants of sigma field can be further obtained from the values of these above moments.

Note that numerically solving Eq.(1) also needs to input the space-time
information of the local temperature and chemical potential, $T (t,x,y,z)$ and $\mu(t,x,y,z)$, for the
effective potential, which are in principle provided by the evolution of a back-ground heat bath.
For simplicity, we assume that the heat bath evolves along simple
trajectories with constant chemical potential (traj. I and traj. II in Fig.~1), and
the temperature drops down in a Hubble-like way~\cite{Mukherjee:2015swa}:
\begin{equation}
\frac{T\left( t\right) }{T_{0}}=\left( \frac{t}{t_{0}}\right) ^{-0.45},
\end{equation}%
where $T_{0}$ is the initial temperature and $t_{0}$ is the initial time.
Considering that the dynamical evolution of the $\sigma $ field belongs
to the universality class of model A, we set the damping
coefficient $\eta$ to be a constant value.

\section{Numerical results}

\begin{figure*}[tbp]
\center
\includegraphics[width=2.9 in, height=2.5in]{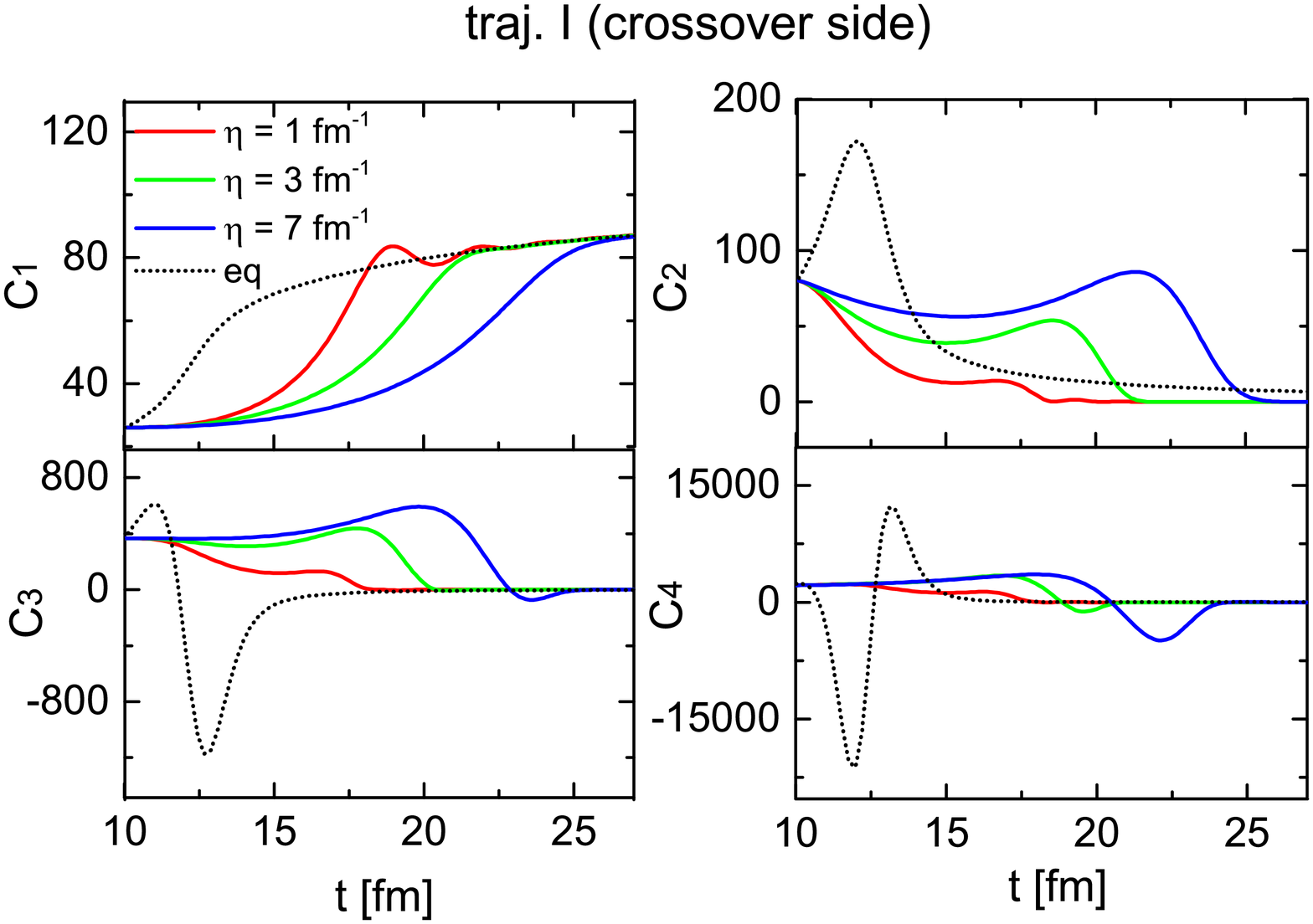}
\includegraphics[width=2.9 in, height=2.5in]{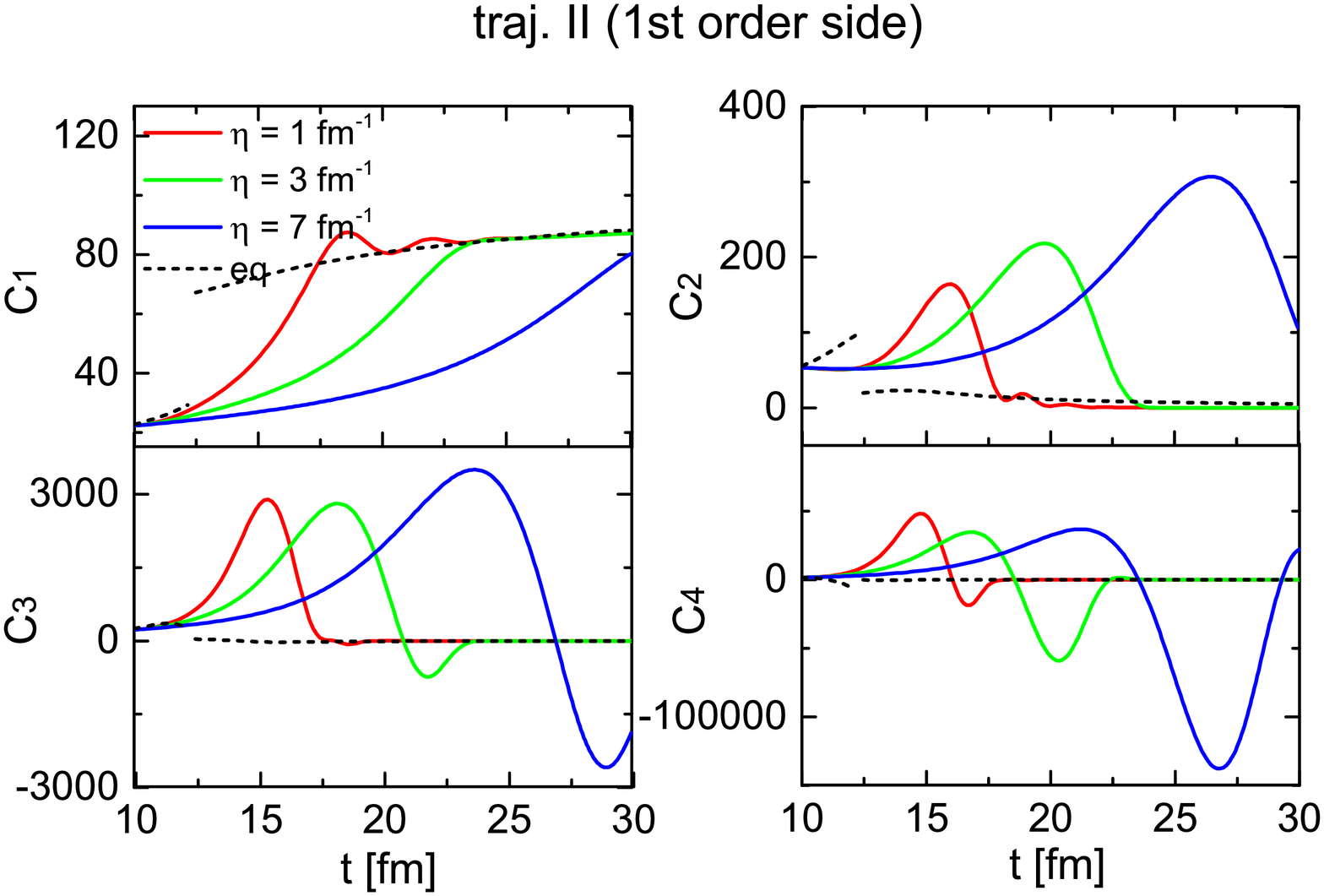}
\vspace{-0.5cm}
\caption{ Dynamical evolution for the cumulants of sigma fields. The left and right panels are the results evolving on the crossover side with $\mu = 200$ MeV (along traj. I) and on the 1st order phase transition side with $\mu = 240$ MeV (along traj. II).}
\label{IPO=1}
\end{figure*}

Fig.~\ref{IPO=1} presents the time evolution for the cumulants of sigma fields. The left and right panels show the results of evolution on the crossover phase transition side (along traj.~I with $\mu = 200$ MeV, which is also close to the critical point) and on the 1st order phase transition side (along traj.~II with $\mu = 240$ MeV), respectively. For each case, we choose
three constant damping coefficients for the dynamical evolution, which are shown as three colored solid lines.  We also plot the equilibrated values of the sigma field (dotted lines) from  the equilibrium critical fluctuations  along traj.~I and traj.~II, using the mapping between temperature and evolution time in Eq.~(5).

For the case with traj.~I, the evolution of the cumulants for critical fluctuations presents clear memory effects. For $C_3$ and $C_4$, the signs and values are different from the equilibrated ones at later evolution time. For example, at t=12 fm/c, both $C_3$ and $C_4$ show a positive sign, which is opposite to the sign of the equilibrated one. In dynamical evolution scenario,  the increase of cumulants  is also delayed due to the critical slowing down.  With the increase of the damping coefficient $\eta$,
the dynamical evolution becomes slower, and behaves like diffusion process. In the early paper~\cite{Jiang:2015hri}, it was found that the equilibrium critical fluctuations always over-predict $C_{2}$ and $C_{3}$ due to the intrinsic positive contributions. The calculations presented in Fig.~2 (left) show that the dynamical evolution  of the sigma field near the critical point could change the sign of $C_3$ and largely delay the increase of $C_2$, which has the potential to qualitatively describe different cumulant data with a properly chosen freeze-out scheme and well tuned parameters.

The right panel presents the dynamical evolution along traj.~II, which is across the
first order phase transition boundary. For the equilibrium values, $C_1-C_4$ show discontinuity  at
the phase transition temperature. Note that the thermodynamical potential has two minima around $T_c$ (Fig.~1, right), which leads to the discontinuity of $C_1-C_4$. For the dynamical evolution scenario,
$C_1-C_4$ continuously change during the evolution and the values of $C_2-C_4$  at late time are much larger than the maximum values of the equilibrated ones. As shown in Fig.~1 (right), there exists a barrier between two minima of the thermodynamical potential, which prevents part of the sigma's configurations evolve
to the real minimum at certain temperatures close to the phase transition.  Such supercooling effect leads the sigma field to be widely distributed in the thermodynamical potential, which also largely
enhances the cumulants  $C_3$ and $C_4$ at the first order transition side.

\section{Summary}
Using Langevin dynamics, we simulate the dynamical evolution of the sigma field
with the effective potential from the linear sigma model.  We found, as
the system evolves in the critical regime,  the memory effects keep
the signs of $C_{3}$ and $C_{4}$ from the early evolution, which
are different from the equilibrated ones at the possible freeze-out points
below $T_{c}$.  For the dynamical evolution across the 1st order phase transition boundary,
the supercooling effect leads the sigma field to be widely distributed in the thermodynamical potential, which largely
enhances the cumulants  $C_3$ and $C_4$, correspondingly.

\section*{Acknowledgements}
We thanks the discussions from U. Heinz, Y. X. Liu, S.~Mukherjee,  M.~Stephanov, D. Teaney ,  H. Stoecker, Y. Yin and
 J. Zheng. This work is supported by the NSFC and the MOST under grant Nos.11435001, 11675004 and 2015CB856900.







\end{document}